\documentclass[10pt, a4paper]{article}
\usepackage[utf8]{inputenc}
\usepackage[final]{lrec2026} 
\usepackage{comment}
\usepackage{soul} 
\usepackage[most]{tcolorbox}
\usepackage{xcolor}
\usepackage{booktabs}
\usepackage{pifont}
\usepackage{float}
\usepackage{listingsutf8}
\usepackage[most]{tcolorbox}
\usepackage{xcolor}
\definecolor{sweqablue}{RGB}{51,102,170}
\definecolor{sweqagreen}{RGB}{76,153,102}
\definecolor{sweqaorange}{RGB}{230,159,0}
\definecolor{sweqared}{RGB}{204,51,51}

\newtcblisting{promptbox}[1][]{
  colback=gray!5,
  colframe=black!60,
  boxrule=0.6pt,
  arc=4pt,
  fonttitle=\bfseries,
  title=#1,
  listing only,
  breakable,
  listing options={
    basicstyle=\ttfamily\footnotesize,
    breaklines=true,
    breakatwhitespace=true,
    columns=fullflexible,
    keepspaces=true,
  }
}

\title{SWE-QA: A Dataset and Benchmark for Complex Code Understanding}

\name{Laïla Elkoussy, Julien Perez$^{\ast,\dagger}$}

\address{
LRE, EPITA*\\
\texttt{laila.elkoussy@epita.fr} \\ \\
$^{\dagger}$Bpifrance \\
\texttt{julien.perez@bpifrance.fr}
}

\abstract{
In this paper, we introduce SWE-QA, a text and code corpus aimed at benchmarking multi-hop code comprehension, addressing the gap between simplified evaluation tasks and the complex reasoning required in real-world software development. 
While existing code understanding benchmarks focus on isolated snippets, developers must routinely connect information across multiple dispersed code segments. 
The dataset comprises 9,072 multiple-choice questions systematically generated from 12 Python repositories of SWE-bench, evaluating several recurrent reasoning patterns like Declaration-and-Call questions that link entity definitions to their usage, and Interacting-Entity questions that examine the dynamic relationships among multiple collaborating components. 
Generated through parsing-based entity extraction and  Large Language Model assisted question construction with carefully validated distractors, the benchmark distinguishes genuine comprehension from superficial pattern matching. 
Evaluation of 15 language models (360M to 671B parameters) reveals significant challenges in multi-hop reasoning, with best performance reaching 74.41\% accuracy. 
Dense architectures consistently outperform mixture-of-experts models by 10-14 percentage points, while reasoning-enhanced variants show inconsistent benefits. 
\\ \newline  }

\begin{document}

\maketitleabstract

\begin{figure*}[ht!]
\centering
\includegraphics[width=\textwidth]{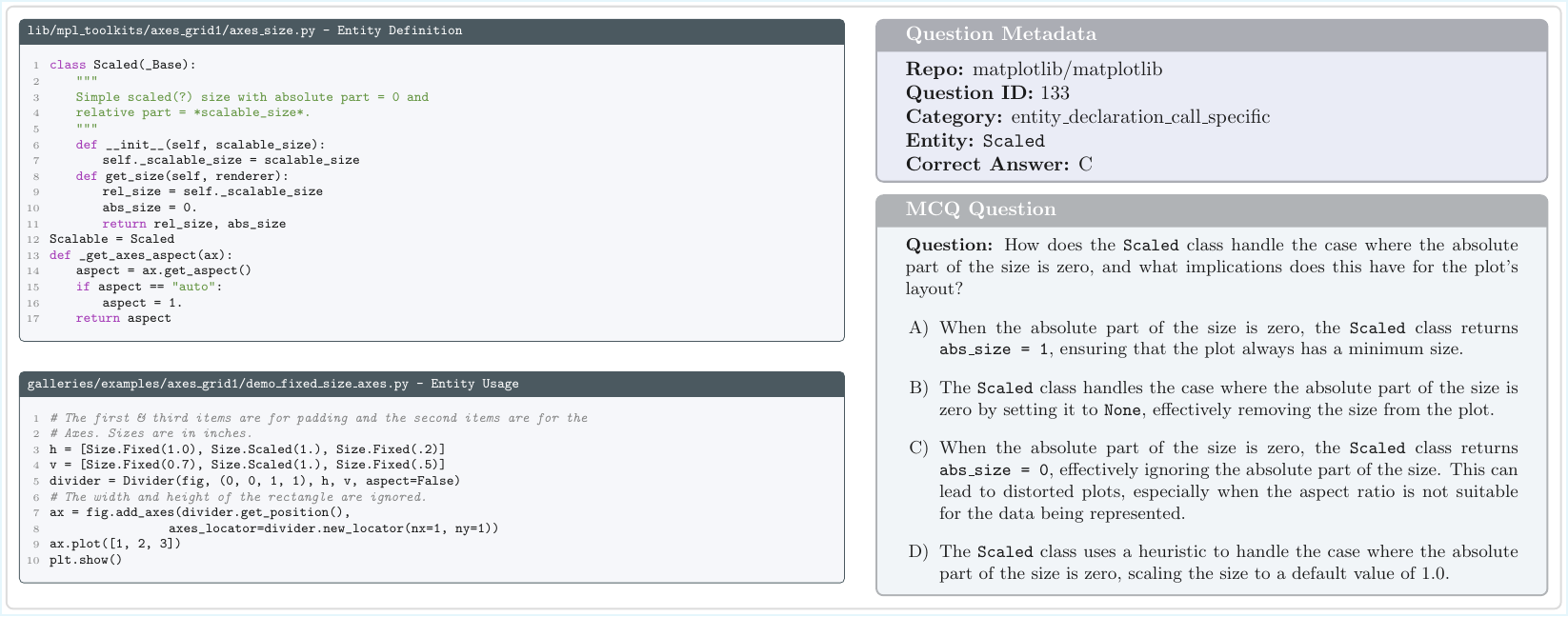}

\caption{Multi-hop question sampled from SWE-QA requiring cross-file reasoning: the \texttt{Scaled} class in \texttt{axes\_size.py} (top left) and its use in \texttt{demo\_fixed\_size\_axes.py} (bottom left). A correct answer must trace how \texttt{abs\_size = 0} in \texttt{get\_size} affects layout when \texttt{Size.Scaled(1.)} is invoked. The multiple-choice design uses targeted distractors to distinguish true cross-context understanding from superficial pattern matching, mirroring the multi-step reasoning developers perform in real codebases.
}

\label{fig:example_1}
\end{figure*}

\section{Introduction}

Code comprehension remains a fundamental challenge in natural language processing for software engineering, with implications extending from academic research to practical software development and automated programming assistance. While significant progress has been made in developing language models for code understanding, capturing syntax, static semantics, and even dynamic behavior \cite{Ma2023LMsSyntaxSemantics}, existing evaluation frameworks predominantly focus on localized reasoning tasks. Such frameworks often fail to capture the complex, project-scale, interconnected nature of real-world software systems, where multi-hop reasoning, interprocedural dependencies, and runtime behavior matter deeply, which is essential for code development agents \cite{She2023Pitfalls, Zhang2023Unifying}.

Current benchmarks typically evaluate models on isolated code snippets or self-contained problems where all necessary information is available within a single context. This misrepresents how software engineers interact with large codebases. Real-world software understanding requires reasoning across multiple, disparate code segments, a process similar to multi-hop reasoning in reading comprehension tasks like HotpotQA \citep{yang2018hotpotqa}, but with added complexity from syntax, execution semantics, and dependencies.

Consider a typical developer workflow: encountering a function call in one file, navigating to its definition in another file to understand parameters and behavior, then tracing how returned values are used elsewhere across modules. This requires synthesizing information from multiple sources, understanding temporal dependencies, and reasoning about data flow patterns, skills that current benchmarks largely fail to assess. Multi-hop code comprehension arises in linking function declarations to invocations across files, tracing class hierarchies, following data pipelines, understanding event-driven architectures, and comprehending complex object interactions.

Despite its importance, existing datasets provide insufficient coverage. Benchmarks like CodeQA, CS1QA, and CodeXGLUE primarily evaluate single-hop reasoning within contained contexts. Even sophisticated benchmarks like SWE-bench, while operating on real repositories, focus on single-issue resolution rather than explicit multi-entity tracking. This gap limits assessment of language models for real-world software engineering. Multi-hop reasoning over large codebases also highlights a limitation of current large language models: finite context windows. Even with extended context, models often struggle to maintain coherence when given excessive or poorly scoped information \citep{liu2023lost, xiao2024scalingcontext}. This underscores the need for evaluation settings that control the amount and structure of context, ensuring performance reflects reasoning ability rather than context length.

We address this gap by introducing a novel dataset and benchmark\footnote{Dataset available at: \url{https://github.com/lailanelkoussy/swe-qa}} designed to evaluate multi-hop, repository-scale code comprehension. Our approach systematically constructs questions requiring understanding of complex relationships between code entities across different file segments, mimicking the reasoning processes developers use in large-scale software systems.
We focus on two fundamental categories: declaration-and-call relationships connecting entity definitions with usage contexts, and interacting entity relationships requiring understanding of how multiple components collaborate. By examining multi-hop scenarios from real repositories, we bridge the gap between simplistic evaluation tasks and real software comprehension, offering an authentic assessment of the cognitive challenges faced by developers and automated systems.

Finally, we leverage a systematic methodology for generating authentic multi-hop code comprehension questions from real software repositories, ensuring relevance, quality, and diversity. Based on this, we curate a dataset with two categories of multi-hop questions and carefully designed distractors, providing a challenging benchmark that tests true understanding rather than memorization. We evaluate multiple state-of-the-art language models on this benchmark, revealing significant performance gaps and highlighting strengths and weaknesses in handling complex code relationships, with attention to scaling effects on multi-hop reasoning. We analyze common failure patterns and error types, offering actionable insights for improving model design and training strategies. Finally, we explore the relationship between model size, architecture, and reasoning performance, providing guidance for practitioners optimizing language models for complex software engineering tasks.

Our contributions are threefold. First, we present SWE-QA, a curated dataset with diverse, high-quality multi-hop questions and carefully designed distractors, enabling robust assessment of genuine code understanding beyond superficial pattern matching. Second, we introduce a systematic methodology for generating this dataset from real repositories, ensuring practical relevance. Third, we evaluate state-of-the-art language models on this benchmark, revealing performance gaps, common failure patterns, and insights for improving multi-hop reasoning over code.

\section{Related Work}

Code comprehension research has primarily focused on single-hop reasoning within isolated code snippets or simple queries. CodeQA \citep{liu-etal-2021-codeqa} provides Java and Python Q\&A pairs from snippets, while CS1QA \citep{sohn-etal-2022-cs1qa} collects Q\&A from introductory programming courses. These datasets emphasize local understanding without requiring reasoning across dispersed code elements.
CRUXEval \citep{gu2024cruxeval} contains 800800
800 short Python functions for assessing reasoning, understanding, and execution through two tasks: CRUXEval-I (input prediction) and CRUXEval-O (output prediction). Each function, generated with Code Llama 34B and filtered for human solvability within a minute, includes input-output examples. Although it tests execution tracing more deeply than prior benchmarks, it remains limited to isolated function-level reasoning.

Code World Models \citep{faircodegenteam2025cwmopenweightsllmresearch} train language models on observation-action trajectories from Python environments to improve execution understanding through world modeling, yet they do not address tracing dependencies across multiple code segments in large repositories.
Broader benchmarks such as CodeXGLUE \citep{lu2021codexglue}, HumanEval \citep{chen2021evaluating}, MBPP \citep{austin2021program}, and SWE-bench \citep{jimenez2024swebench} evaluate realistic programming tasks like completion, translation, and patch generation, but still lack multi-element reasoning. LocAgent \citep{chen2025locagent} addresses the related problem of code localization---identifying where in a codebase changes must be made---by parsing repositories into directed heterogeneous graphs that capture files, classes, functions, and their dependencies (imports, invocations, inheritance), and leveraging LLM agents to navigate these graphs via multi-hop reasoning. While LocAgent demonstrates strong performance on file-level localization and downstream issue resolution, it is task-specific and does not provide a comprehension benchmark for evaluating cross-context understanding across diverse reasoning patterns.
In natural language processing, multi-hop datasets such as HotpotQA \citep{yang2018hotpotqa}, MuSiQue \citep{trivedi-etal-2022-musique}, WikiHop, and WikiMultihopQA \citep{welbl-etal-2018-constructing} combine information across sources. Applying this to code is difficult due to syntax, dependencies, and execution semantics. Table \cite{tab:benchmark-comparison} in Appendix ~\ref{appendix:comparison-benchmarks}  summarizes the key dimensions along which existing benchmarks differ, highlighting the gap that SWE-QA is designed to fill. Our work bridges this gap by building a multi-hop code comprehension dataset that extends multi-step reasoning to software repositories, evaluating models on tracing logical dependencies and entity interactions across repository-scale codebases.

\begin{figure}[!ht]
\begin{center}
\includegraphics[width=\columnwidth]{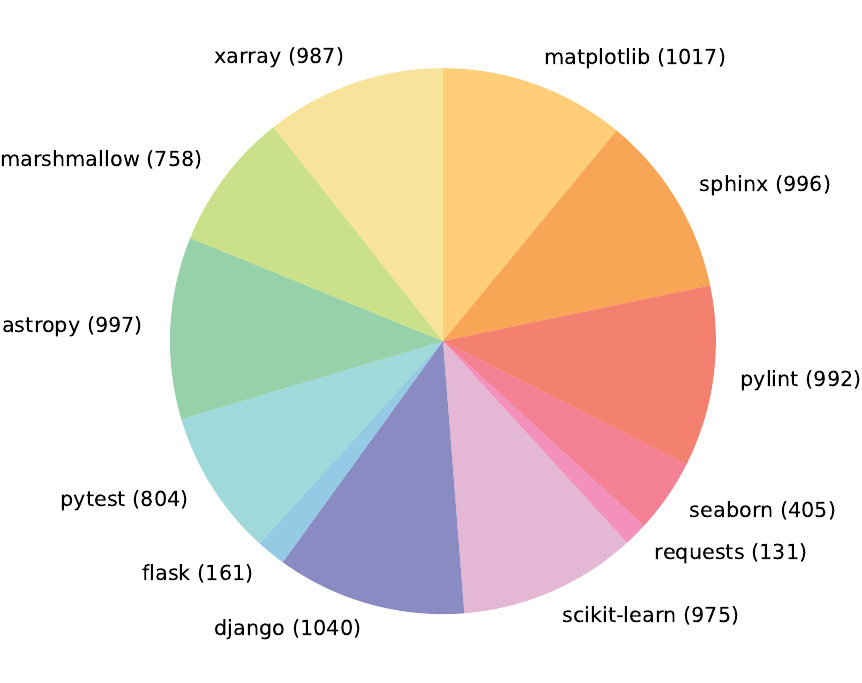}
\caption{Distribution of questions of SWE-QA across public repositories. The repositories are 12 open source GitHub repositories that each contains the source code for a popular, widely downloaded PyPI package.}
\label{fig:repo_distribution}
\end{center}
\end{figure}

\section{Methodology}

Our approach comprises three steps: code processing and chunking, multi-hop question generation, and quality control through post-processing and benchmarking. This pipeline produces diverse, high-quality questions that test complex code understanding in realistic software engineering contexts.

\subsection{Code Processing and Chunking}

\paragraph{Repository Selection.} We select 12 Python repositories from the SWE-bench dataset to cover diverse domains, as web frameworks, data processing, scientific computing, utilities, and varying code complexity. Preliminary analysis of entity density, file structure, and cross-file dependencies informed chunking parameters.

\paragraph{Text Segmentation.} We used LangChain's recursive character splitter with Python-aware separators to preserve syntactic boundaries. Chunks of 1000 characters with zero overlap balance context and efficiency, typically keeping functions or classes intact. Separators follow a hierarchy: double newlines, class/function definitions, control structures, single newlines with indentation, and commas. Boundaries preserve syntactic integrity to avoid broken statements.

\paragraph{Entity Extraction.}
We built an entity extraction pipeline based on Abstract Syntax Trees (AST) to systematically analyze Python source code and identify structural and semantic elements. Using Python’s built-in AST parser, the system detects declared entities such as classes, functions, methods, variables, and constants, capturing attributes including type, scope, inferred data type, and defining context. Function and method invocations are recorded separately as called entities, distinguishing definitions from usages.

This dual representation enables reconstruction of call graphs, inheritance hierarchies, and data dependencies. Extracted entities are mapped back to their corresponding code chunks through structural and lexical matching, allowing each chunk to reference the entities it declares and calls. The resulting bidirectional mapping links entities and chunks even when definitions and usages span multiple segments, forming a semantic graph that supports higher-level analyses of dependency and modular structure across the codebase.

\begin{figure}[!ht]
\begin{center}
\includegraphics[width=\columnwidth]{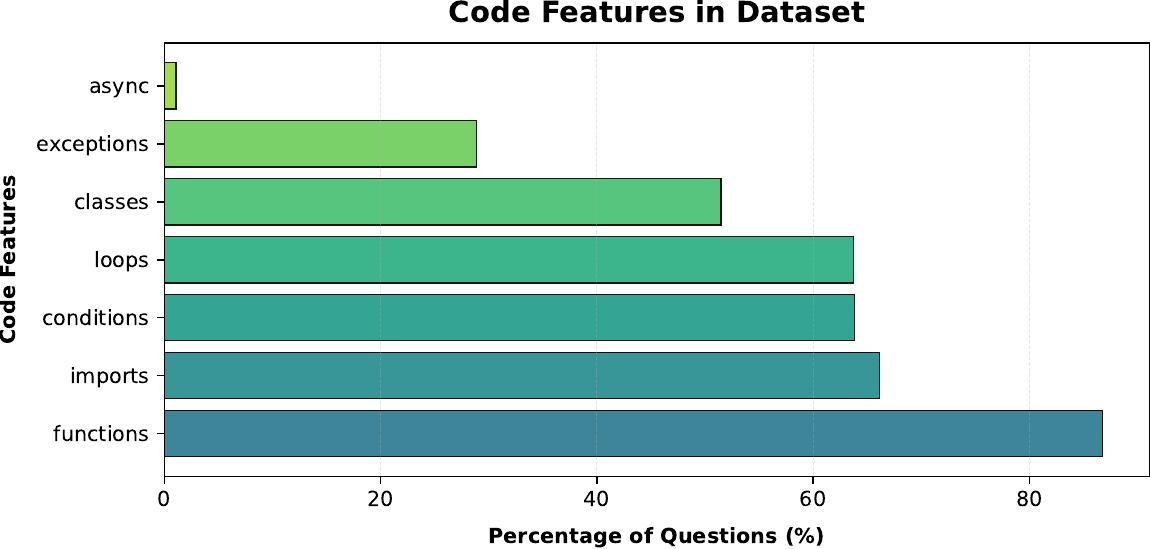}
\caption{Ratios of code chunks showing the presence of specific programming constructs (loops, conditions, functions, classes, imports, async operations, and exceptions).}
\label{fig:code_features}
\end{center}
\end{figure}

\subsection{Question Categories and Generation}

\paragraph{Multi-Hop Taxonomy.} Inspired by the HotpotQA dataset, we have defined two multi-hop question types. Declaration-and-Call (DC) questions connect an entity’s definition in one chunk with its usage in another, requiring reasoning about parameters, return values, state changes, or error handling. Interacting Entity (IE) questions involve three chunks where two entities interact, demanding analysis of data flow, execution order, shared state, or collaborative behaviors.

\paragraph{Chunk Sampling.} Candidate pairs for DC and triplets for IE were enumerated across repositories and validated for genuine multi-hop relationships. Selection enforces diversity: limiting questions per entity or entity pair, balancing repositories by size and complexity, and ensuring a mix of simple and complex patterns. We capped each repository at 600 questions per category and applied quality thresholds for relationship strength and code complexity.

\subsection{Question and Answer Generation}

\paragraph{Question Generation Process.} We used \texttt{Meta-Llama/Llama-3.2-3B-Instruct}, tuned for creative but precise output, with prompts specifying context, question type, and multi-hop reasoning requirements. An iterative refinement checked clarity and multi-hop validity. Semantic analysis ensured structural diversity, varied vocabulary, and appropriate complexity.

\begin{figure}[!ht]
\begin{center}
\includegraphics[width=\columnwidth]{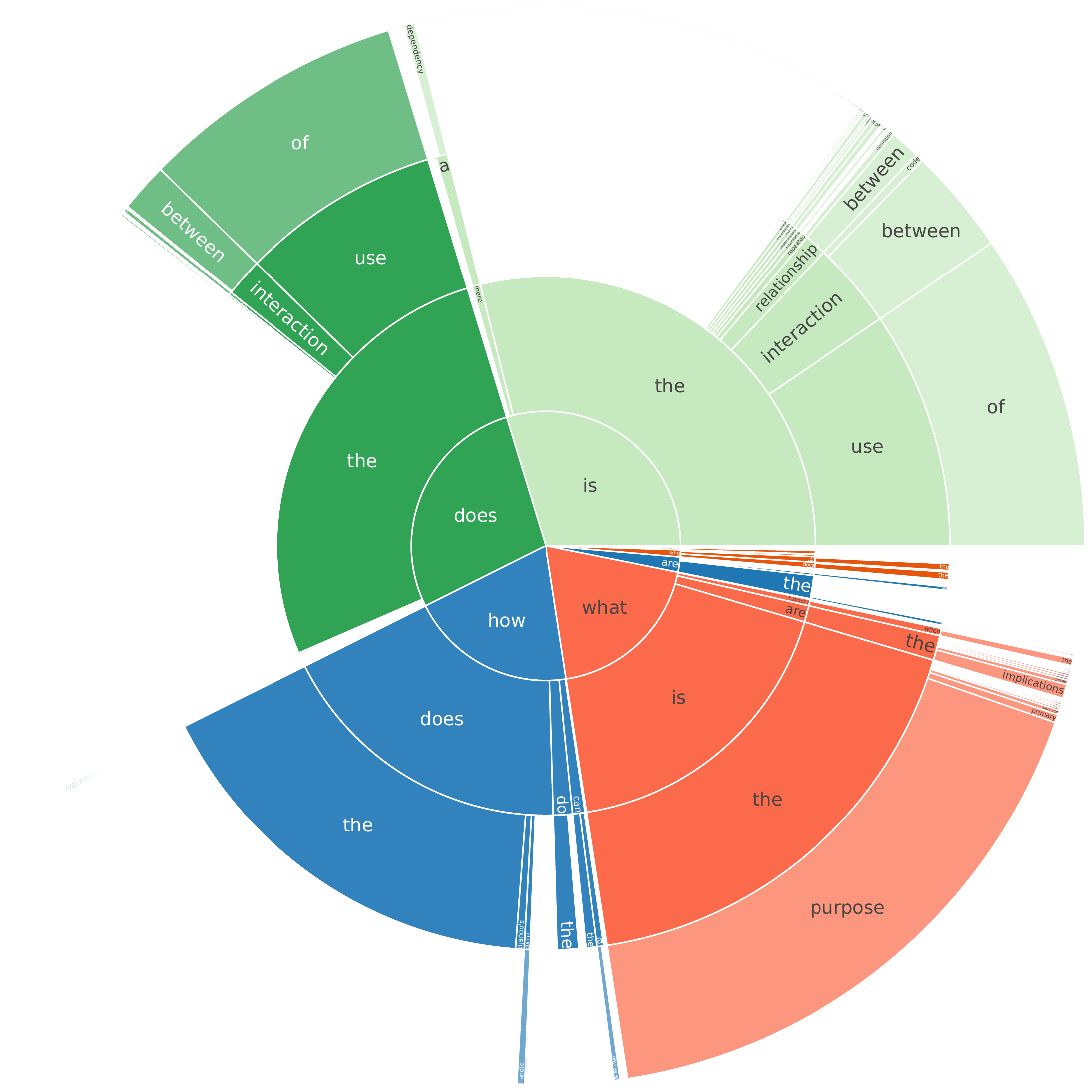}
\caption{Distribution of the first four words of all questions, representing their frequency patterns. Empty
colored blocks indicate suffixes that are too rare to
show individually.}
\label{fig:questions_sunburst}
\end{center}
\end{figure}

\paragraph{Answer Generation.} Answers were generated by the same model used for question creation, conditioned on the question and relevant code chunks, then refined for conciseness through a follow-up call. Prompt validity was verified by human review on a subset before scaling to the full dataset.

\paragraph{Distractor Creation and Stylistic Alignment.} Three distractors per question were generated in a single call conditioned on the question, code, and correct answer, ensuring technical plausibility while reflecting common misconceptions. After validating format compliance and cardinality, correct answers were adapted to match the linguistic style of distractors, preventing surface-level pattern matching. All generation and adaptation used the same model, with human-verified prompts guiding each stage.

\section{Experimental Setup}
\label{sec:experimental-setup}

We evaluate a collection of language models on SWE-QA to assess their multi-hop code comprehension capabilities and the pertinence of our corpus. 
The evaluated models span from small to large scales and include two \texttt{SmolLM2} variants, \texttt{Llama-3.3-70B-Instruct}, and \texttt{DeepSeek-R1}. 
To isolate the effect of reasoning on multi-hop question answering, we include both reasoning and non-reasoning variants of the same models, such as \texttt{Phi-4-mini} and \texttt{Qwen3-4B}.
All evaluations are conducted in a zero-shot setting: models receive only the relevant code chunks, the question, and the multiple-choice options—without any additional in-context examples. 
SWE-QA comprises \textbf{9,072} questions drawn, including \textbf{4,584} DC questions and \textbf{4,488} IE questions. We report overall accuracy, category-specific accuracy, repository-level performance, and retrieval quality metrics such as NDCG@$k$ and Precision@$k$, where applicable. 
Post-processing scripts normalize model outputs to extract the selected option, addressing cases where responses include explanations or formatting variations.

\paragraph{Oracle Question Answering.}
In the first setting, models are evaluated with an \emph{oracle retrieval}. For each question, the code chunks that are relevant are solely provided, without any distractors. Models must return a single answer based on these oracle-provided chunks, the question, and the multiple-choice options. This setting establishes an upper bound on comprehension performance by removing retrieval errors.

\paragraph{Retrieval-Based Evaluation.}
The second setting evaluates retrieval performance. All code chunks of each repository are embedded using the \texttt{Salesforce/SFR-Embedding-Code-400M\_R} \cite{liu2025codexembedgeneralistembeddingmodel} model which is optimized for code/text retrieval. 
For each question, a maximum inner product search is performed on the repository’s vector database to retrieve the ten most relevant chunks.
Retrieval quality is measured using NDCG@k and Precision@k, providing insight into the difficulty of locating relevant information compared to the ideal oracle scenario.

\paragraph{Noisy Oracle Comprehension.}
Finally, we test the best-performing models under a \emph{noisy oracle} setting. 
Each question is presented with the relevant chunks plus a set of distractor chunks. These distractors are selected based on the most likely but irrelevant chunks retrieved in the second experiment. 
This setting evaluates whether models can still leverage the correct answer when the context is contaminated with retrieval-induced noise, better reflecting real-world code understanding pipelines.

\paragraph{Dataset Validation and Correction}
LLM-based answer generation occasionally produces incorrect labels. To improve label quality, we employed consensus-based validation on the oracle question answering results. We excluded three small models (SmolLM2-360M, SmolLM2-1.7B, DeepSeek-R1-Distill-Qwen-1.5B) to prevent capacity-related noise from corrupting consensus signals. Using two criteria: fewer than 3 models selecting the generated answer, and at least 11 of 12 retained models agreeing on an alternative. This identified 66 candidate mislabeled questions. Spot-check validation on 10 randomly sampled questions confirmed that the consensus answer was correct in all cases (100\% precision), justifying batch correction of all 66 questions. All results reported below reflect corrected ground truth labels, ensuring performance metrics reflect genuine code comprehension rather than label artifacts.

\section{Experiments}
\label{sec:results}

All results reported below reflect corrected ground truth labels after applying our validation procedure described in Section \ref{sec:experimental-setup}. 
\begin{figure}[!ht]
\begin{center}
\includegraphics[width=\columnwidth]{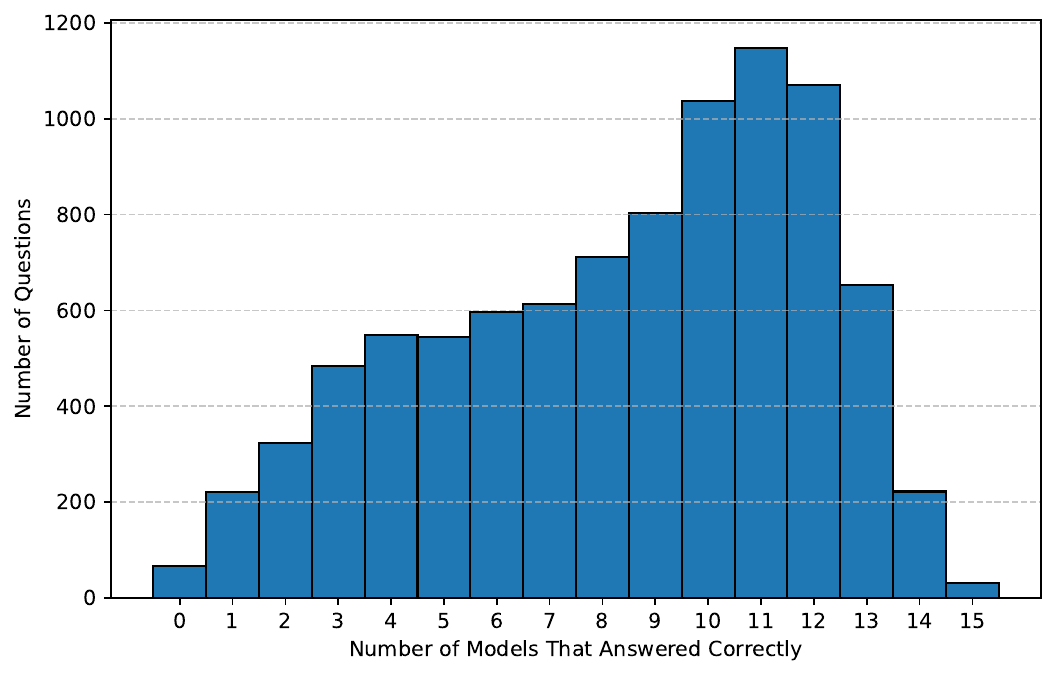}
\caption{Histogram showing question difficulty, measured by the number of models that answered each question correctly. Fewer correct responses indicate higher difficulty.}
\label{fig:question_difficulty_histogram}
\end{center}
\end{figure}
\begin{figure*}[ht!]
\centering
\includegraphics[width=\textwidth]{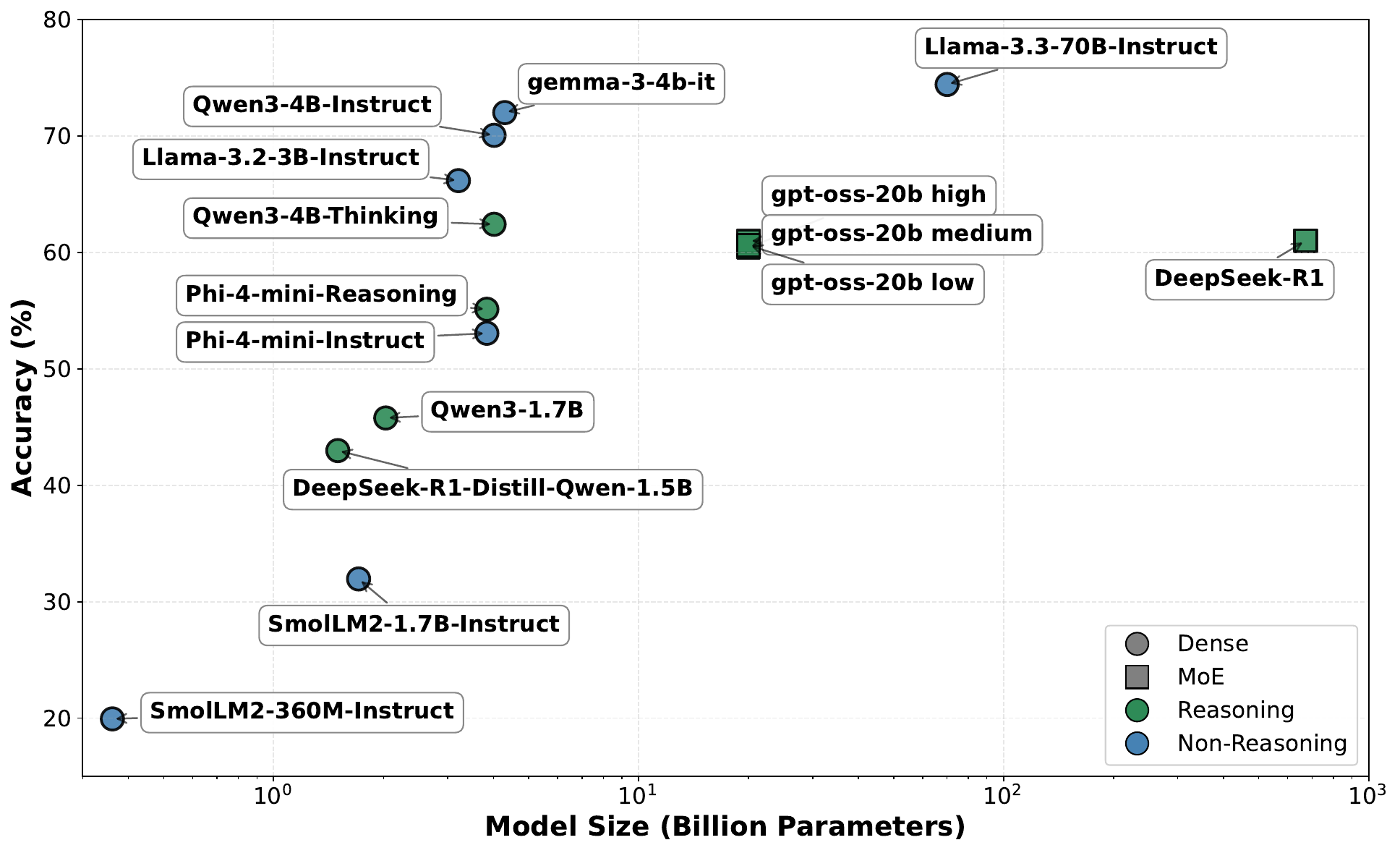}
\caption{Model accuracy in the Oracle Question Answering setting as a function of parameter count.
Circle markers denote dense architectures, squares indicate MoE models.
Green corresponds to reasoning-tuned models, while blue represents standard instruction-tuned models.}
\label{fig:model_performance}
\end{figure*}
\subsection{Oracle Question Answering}
\paragraph{Overall Performance.}
Across 15 models, accuracy spans 74.41\%–19.94\%, highlighting the difficulty of multi-hop code reasoning. \texttt{Llama-3.3-70B-Instruct} leads with 74.41\%, 2.4 points above \texttt{gemma-3-4b-it}, demonstrating that optimized smaller models can approach large-scale performance. \texttt{Qwen3-4B-Instruct} follows at 70.05\%, while \texttt{Llama-3.2-3B-Instruct} shows rare balance across question types, hinting at architecture-driven inter-entity strengths. Figure \ref{fig:question_difficulty_histogram} shows a broad distribution of question difficulty: nearly half of the questions are correctly answered by at least 10 models, while roughly one-tenth are solved by at most three, providing a meaningful spectrum to differentiate model capabilities.

\paragraph{Architecture and Reasoning Effects.}
The two MoE model families evaluated underperform their dense counterparts across scales. \texttt{gpt-oss-20b} variants cluster around 60\% accuracy, 10 points below dense models like \texttt{gemma-3-4b-it}, with negligible variation across reasoning depths. \texttt{DeepSeek-R1}, despite 671B total and 37B active parameters, achieves only 60.98\%, suggesting a potential scaling plateau not observed in the dense models evaluated. One possible explanation is that multi-hop reasoning demands unified knowledge access, while MoE routing may fragment information across expert boundaries, making cross-context integration more difficult.

Reasoning-enhanced variants show inconsistent benefits. \texttt{Qwen3-4B-Thinking} underperforms its instruct counterpart by 7.6 points, particularly on IE questions, indicating that extended reasoning at small scales may amplify error propagation. Phi-4-Mini gains only two points from reasoning, showing limited benefit under capacity constraints. Based on the models evaluated, neither extended reasoning nor larger MoE scale appears to consistently improve multi-hop comprehension, though broader evaluation across more model families would be needed to substantiate this observation.
\begin{table}
\centering
\small 
\begin{tabular}{lcc}
\hline
\textbf{Model} & \textbf{DC (\%)} & \textbf{IE (\%)} \\
\hline
\texttt{\textbf{Llama-3.3-70B-Instruct}} & \textbf{78.75} & \textbf{69.98} \\
\texttt{gemma-3-4b-it} & 75.32 & 68.60 \\
\texttt{Qwen3-4B-Instruct} & 74.58 & 65.41 \\
\texttt{Llama-3.2-3B-Instruct} & 66.79 & 65.50 \\
\texttt{Qwen3-4B-Thinking} & 69.85 & 54.81 \\
\texttt{gpt-oss-20b} (medium) & 67.95 & 53.83 \\
\texttt{DeepSeek-R1} & 68.23 & 53.58 \\
\texttt{gpt-oss-20b} (low) & 67.62 & 53.43 \\
\texttt{gpt-oss-20b} (high) & 67.51 & 53.16 \\
\texttt{Phi-4-Mini-Reasoning} & 59.77 & 50.35 \\
\texttt{Phi-4-Mini-Instruct}& 58.09 & 47.90 \\
\texttt{Qwen3-1.7B} & 48.32 & 43.18 \\
\texttt{DeepSeek-R1-Dist-Qwen-1.5B} & 46.72 & 39.14 \\
\texttt{SmolLM2-1.7B-Instruct} & 33.92 & 29.94 \\
\texttt{SmolLM2-360M-Instruct} & 19.28 & 20.61 \\
\hline
\end{tabular}
\caption{\label{tab:model-scores-cat} Category-level accuracy results for Oracle Question Answering, showing performance on DC and IE questions.}
\end{table}

\begin{table}[!ht]
\centering
\begin{tabular}{llll}
\hline
\textbf{Metric} & \textbf{$k=3$} & \textbf{$k=5$} & \textbf{$k=10$}  \\
\hline
Precision@$k$ & 0.2288 & 0.1626 &  0.0961 \\
Recall@$k$ & 0.3305 & 0.3875 & 0.4535 \\
F1@$k$ & 0.2643 & 0.2248 &  0.1567 \\
Hit Rate@$k$ & 0.5909  & 0.6620 &  0.7341 \\
MRR@$k$ &0.4945 &  0.5108 & 0.5206 \\
NDCG@$k$ &0.3430 & 0.3726 & 0.3998 \\
\hline
\end{tabular}
\caption{\label{retrieval-scores} Retrieval-based evaluation results with $k=3,5,10$.}
\end{table}

\paragraph{Category-Specific Patterns.}
As shown per Table \ref{tab:model-scores-cat}, models consistently perform better on DC than IE questions, with gaps going from 5 to 15 points. Reasoning models such \texttt{Qwen3-4B-Thinking}, \texttt{DeepSeek-R1}, and \texttt{gpt-oss-20b} show the largest disparities, confirming that reasoning over interacting entities across three code locations is substantially harder than single-entity tracing. Top performers such as \texttt{Llama-3.3-70B-Instruct}, \texttt{gemma-3-4b-it}, and \texttt{Qwen3-4B-Instruct} exhibit smaller gaps (6–9 points), while \texttt{Llama-3.2-3B-Instruct} stands out with nearly balanced results, suggesting architectural features that enhance inter-entity reasoning even at small scale.

\begin{table}[!ht]
\begin{center}
\includegraphics[width=\columnwidth]{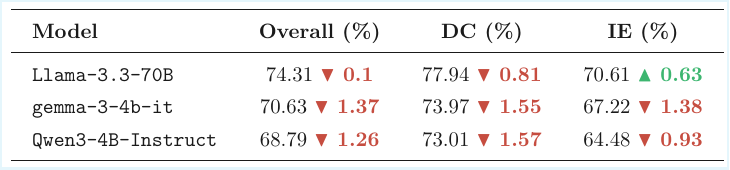}
\caption{Difference of model performance in Noisy Oracle Question Answering compared to Oracle Question Answering}
\label{table:model_performance_distractors}
\end{center}
\end{table}

\paragraph{Model Size vs. Performance Analysis.}
Performance scales with size but is conditioned by architecture and training methods. \texttt{Llama-3.3-70B-Instruct} leads, confirming benefits from scale, though \texttt{gemma-3-4b-it}’s near parity shows the power of optimization. \texttt{DeepSeek-R1}’s weak result despite 37B active parameters shows that scale alone is insufficient. The 3–4B range emerges as a performance “sweet spot,” showing the widest variance driven by architecture and inference strategy. Below 2B parameters, accuracy collapses indicating a lower bound near 3B parameters of current models for effective multi-hop reasoning.

\subsection{Retrieval-Based Evaluation}
Table~\ref{retrieval-scores} shows moderate retrieval effectiveness. Precision@$k$ declines from 0.23 ($k=3$) to 0.10 ($k=10$), while Recall@$k$ increases from 0.33 to 0.45, reflecting the typical precision-recall trade-off. The highest F1@$k$ of 0.26 at $k=3$ indicates smaller retrieval sets better balance relevance and noise. Hit Rate@$k$ above 0.59 and stable MRR@$k$ around 0.5 show relevant chunks frequently appear in top ranks, though NDCG@$k$ (0.34–0.40) suggests limited ranking quality. Retrieval is sufficient for downstream tasks but leaves room for improvement.
\begin{table*}[ht!]
\centering
\includegraphics[width=\textwidth]{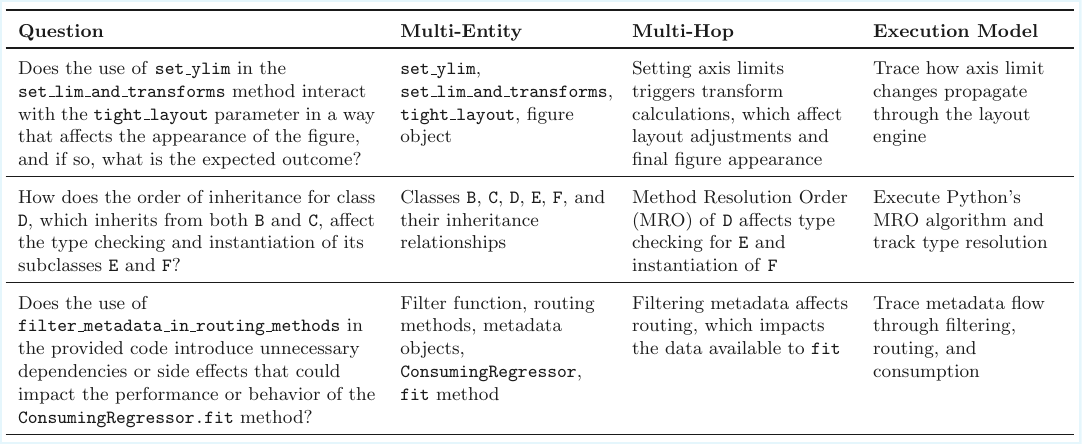}
\caption{Examples of challenging questions requiring combinations of reasoning capabilities. Questions shown were answered correctly by at most two models, both top performers, demonstrating the difficulty frontier of current multi-hop code comprehension.}
\label{tab:reasoning_examples}
\end{table*}

\subsection{Distractor and Oracle Comprehension}
Table~\ref{table:model_performance_distractors} reveals distinct robustness patterns under retrieval noise. \texttt{Llama-3.3-70B-Instruct} shows negligible degradation (0.04-point drop), with slight DC decline but improved IE performance, suggesting noise can aid distractor elimination. \texttt{gemma-3-4b-it} degrades modestly ($\sim$1 point) across categories, while \texttt{Qwen3-4B-Instruct} exhibits the largest decline yet retains strong accuracy. Noise tolerance scales with model size, though architecture also matters.

\section{Discussion}

\subsection{Architectural Insights}
\label{subsec:architectural-insights}

\paragraph{MoE vs. Dense Models.}
In our experiments, the two MoE model families tested underperform their dense counterparts, potentially pointing to structural limitations in multi-hop reasoning, though caution is warranted given the small sample. \texttt{DeepSeek-R1} achieves only 60.98\%, similar to \texttt{gpt-oss-20b} and below dense counterparts like \texttt{Llama-3.3-70B-Instruct} and \texttt{gemma-3-4b-it}. This parity across 4–37B active parameters may suggest a scaling plateau not observed in dense architectures, though differences in training data and optimization could equally account for the gap. Possible contributing factors include fragmented expert knowledge hindering cross-context integration, routing difficulties in activating coherent expert sets, or scaling inefficiency. The wider DC–IE gaps of approximately 14-15 pts in MoE models compared to 6–9 pts in dense models are consistent with this pattern, though they do not establish a causal architectural explanation.

\paragraph{Reasoning-Enhanced Inference.}
Reasoning capacity has not yet shown consistent benefits for multi-hop code comprehension.  \texttt{Qwen3-4B-Thinking} underperforms its non-reasoning counterpart by 7.6 points, \texttt{Phi-4-Mini} gains only marginally, and \texttt{gpt-oss-20b} exhibits near-identical outcomes across reasoning intensities. These results indicate that extended reasoning, regardless of depth, does not reliably improve performance on this task.

\subsection{Qualitative Analysis of Challenges}

To better understand what makes certain questions particularly challenging, we conducted a qualitative analysis of failure patterns. We identify three primary axes of difficulty that characterize the reasoning requirements in multi-hop code comprehension. First, multi-entity reasoning refers to questions requiring simultaneous tracking of multiple interacting entities (classes, functions, variables) and their relationships across code segments. Second, multi-hop reasoning denotes questions demanding a sequential chain of logical steps connecting information across dispersed code locations, where each step depends on the previous one. Third, execution modeling encompasses questions necessitating mental simulation of code execution to trace value transformations, state changes, or control flow across multiple components.

These axes are not mutually exclusive; the most challenging questions typically combine all three dimensions. To illustrate these reasoning requirements, we analyzed questions that were answered correctly by at most two models, and only by top performers (Llama-3.3-70B-Instruct or gemma-3-4b-it). This selection criterion eliminates questions likely solved by random guessing while focusing on tasks that remain challenging yet theoretically within reach of current capabilities.

Table~\ref{tab:reasoning_examples} presents three representative examples exhibiting distinct patterns of complexity. The first question requires understanding how \texttt{set\_ylim} method calls propagate through transform calculations to affect layout adjustments—exemplifying multi-hop reasoning through a rendering pipeline with moderate entity complexity. The second question demands analysis of Python's Method Resolution Order (MRO) across a five-class inheritance hierarchy, requiring precise execution modeling of type resolution alongside tracking multiple class relationships. The third question combines all three dimensions: tracking metadata flow through filtering and routing functions while modeling side effects on the \texttt{fit} method's behavior.

These examples reveal that model failures often stem not from inability to understand individual code constructs, but from difficulty maintaining coherent reasoning chains across multiple logical steps while simultaneously tracking entity states and simulating execution semantics. The architecture-specific performance gaps observed in our quantitative results (Section \ref{subsec:architectural-insights}) likely reflect differential capabilities along these reasoning axes, with the MoE models in our sample tending to struggle more when entity tracking and execution modeling must be performed jointly, possibly reflecting routing constraints across expert boundaries.

\section{Conclusion}
We introduced SWE-QA, a benchmark for multi-hop code comprehension that evaluates language models on complex reasoning tasks across real software repositories. SWE-QA captures core challenges of large-scale code understanding and mirrors the cognitive demands faced by developers navigating extensive codebases.

The evaluation of fifteen models shows that multi-hop reasoning remains a major obstacle, with persistent weaknesses in handling dispersed and interdependent code contexts. Our analysis highlights three primary axes of difficulty: multi-entity reasoning, multi-hop reasoning, and execution modeling, revealing gaps in current architectures and training approaches.

Future work should expand human annotation to strengthen label reliability, explore systematic prompt design to isolate methodological effects, and extend SWE-QA to additional programming languages. Controlled comparisons across architectures, targeted fine-tuning, and human baselines can further clarify the limits of current models. Explicit modeling of entity relationships and execution dynamics may ultimately enable more structured and execution-aware reasoning.

Overall, this benchmark establishes a foundation for advancing study of complex code comprehension and guiding the development of models with deeper compositional reasoning capabilities.

\section{Limitations}

\paragraph{Dataset Construction.}
SWE-QA is restricted to Python repositories from SWE-bench and to 2--3 hop reasoning chains, limiting generalization to other programming languages and to deeper multi-hop scenarios. Complex code patterns such as asynchronous control flow, metaprogramming, and cross-module dynamic dispatch are largely absent, as they fall outside what static syntactic analysis can reliably model. Questions were generated by \texttt{Llama-3.2-3B-Instruct}; larger generation models may produce higher-quality or more diverse questions. Fixed 1000-character chunking may not optimally suit all code structures or context windows, and training data contamination for the evaluated models cannot be fully excluded.

\paragraph{AST-Based Generation Bias.}
Entity extraction relies on Python's built-in AST parser, which captures only statically analyzable, syntactically explicit relationships. As a result, benchmark questions are inherently grounded in the structural view of code that the AST provides: declared entities, explicit call sites, and syntactic inheritance chains. This introduces a systematic bias: models and systems that navigate code through similar structural or symbolic lenses may be artificially advantaged. In particular, DC questions, which trace a declaration to its call site, directly mirror the entity relationships that AST-based code navigation tools expose. Systems with access to AST-backed tooling, such as language servers or MCP-enabled tools offering \textit{go-to-definition} and \textit{find-references} capabilities, operate on the same representation used to generate the questions and therefore have an inherent structural advantage. Conversely, dynamic code behaviors invisible to static parsing, including runtime polymorphism, decorator-induced modifications, and reflective patterns, are systematically underrepresented. Results should therefore be interpreted with the caveat that the benchmark measures comprehension of statically identifiable code structure, and comparisons involving AST-tool-assisted systems may not reflect genuine differences in code understanding.

\paragraph{Evaluation Protocol.}
All models were evaluated in zero-shot settings without systematic prompt engineering, potentially underestimating models sensitive to prompt format or those that benefit from structured reasoning. Though \texttt{gpt-oss-20b} was tested across thinking intensities, other models used default settings and reasoning hyperparameters were not exhaustively explored. Retrieval used a single embedding model, leaving open whether alternative retrieval strategies would change the relative difficulty of the retrieval-based setting.

\paragraph{Architectural Conclusions.}
MoE underperformance was observed on only two model families, limiting the strength of architectural generalizations. Correlations between architecture and performance do not establish causality: observed differences may reflect training data composition, optimization choices, or scale rather than fundamental architectural constraints. Claims about MoE limitations should therefore be treated as preliminary observations warranting controlled investigation rather than definitive conclusions.

\section{Ethical Considerations}

This work uses publicly available code repositories and focuses on advancing code comprehension capabilities. All code processing and analysis respect the original licenses and usage terms of the source repositories.


\section{Bibliographical References}\label{sec:reference}
\bibliographystyle{lrec2026-natbib}
\bibliography{references} 
\clearpage
\onecolumn
\appendix

\section{Comparison to Existing Benchmarks} \label{appendix:comparison-benchmarks}
\begin{table}[H]
\centering
\footnotesize
\begin{tabular}{l|c|c|c|c|c}
\toprule
\textbf{Benchmark} & \textbf{Task} & \textbf{Scope} & \textbf{Multi-hop} & \textbf{Cross-file} & \textbf{Real Repos} \\
\midrule
CodeQA & QA & Snippet & \textcolor{sweqared}{\ding{55}} & \textcolor{sweqared}{\ding{55}} & \textcolor{sweqared}{\ding{55}} \\
CS1QA & QA & Snippet & \textcolor{sweqared}{\ding{55}} & \textcolor{sweqared}{\ding{55}} & \textcolor{sweqared}{\ding{55}} \\
CRUXEval & I/O Prediction & Function & \textcolor{sweqared}{\ding{55}} & \textcolor{sweqared}{\ding{55}} & \textcolor{sweqared}{\ding{55}} \\
Code World Models & World Modeling & Function & \textcolor{sweqared}{\ding{55}} & \textcolor{sweqared}{\ding{55}} & \textcolor{sweqared}{\ding{55}} \\
CodeXGLUE & Various & File & \textcolor{sweqared}{\ding{55}} & \textcolor{sweqared}{\ding{55}} & Partial \\
HumanEval & Generation & Function & \textcolor{sweqared}{\ding{55}} & \textcolor{sweqared}{\ding{55}} & \textcolor{sweqared}{\ding{55}} \\
MBPP & Generation & Function & \textcolor{sweqared}{\ding{55}} & \textcolor{sweqared}{\ding{55}} & \textcolor{sweqared}{\ding{55}} \\
SWE-bench & Patch Gen. & Repository & Implicit & \textcolor{sweqagreen}{\ding{51}} & \textcolor{sweqagreen}{\ding{51}} \\
LocAgent & Localization & Repository & Implicit & \textcolor{sweqagreen}{\ding{51}} & \textcolor{sweqagreen}{\ding{51}} \\
\midrule
\textbf{SWE-QA (Ours)} & \textbf{MCQ} & \textbf{Repository} & \textcolor{sweqagreen}{\ding{51}} & \textcolor{sweqagreen}{\ding{51}} & \textcolor{sweqagreen}{\ding{51}} \\
\bottomrule
\end{tabular}
\caption{Comparison of existing code benchmarks across key dimensions. SWE-QA is the only benchmark to jointly support explicit multi-hop reasoning, cross-file context, and real repository grounding.}
\label{tab:benchmark-comparison}
\end{table}
\section{Prompt Specifications}

This section documents all prompts used in the \texttt{QuestionMaker} module.
For each prompt, we explicitly distinguish between:

\begin{itemize}
    \item \textbf{System Instructions} — Instructions sent as the system message
    \item \textbf{User Message} — Content sent as the user message
\end{itemize}

Placeholders enclosed in \texttt{\{\}} are programmatically substituted before sending. An explanation of the placeholders is available in Section~\ref{sec:placeholders}. 

\subsection{Entity-Specific Question Generation}

\textbf{Type:} System + User message \\
\textbf{Purpose:} Generate a focused question about a specific entity

\begin{promptbox}[System Instructions]
You will be given one or more code snippets, possibly from multiple files.
A specific entity (such as a class, function, or variable) will be identified.

Entity of Focus: {entity_name}

Task:
- Write one clear and concise question about this entity.
- The question should highlight something a developer might consider,
  such as purpose, behavior, interactions, or improvements.
- Keep the question short and direct.
- Do not explain the code or provide an answer.

Output format:
Question: <your question here>
\end{promptbox}

\subsubsection*{User Message}

\begin{verbatim}
<JOINED CODE CHUNKS CONTAINING ENTITY>
\end{verbatim}

\subsection{Interacting Entities Question Generation}

\textbf{Type:} Single user prompt (no separate system message) \\
\textbf{Purpose:} Generate a question about the interaction between two entities

\subsubsection*{User Prompt Template}

\begin{promptbox}[User Prompt]
You are given two code entities, {entity_A} and {entity_B},
along with a snippet where they interact.

Your task is to write one clear and concise question about their relationship.

Input:
- {entity_A} Definition Code:
{entity_A_definition_code}

- {entity_B} Definition Code:
{entity_B_definition_code}

- Interaction Code:
{entity_interaction_code}

Guidelines:
- Ask about design, abstraction, dependencies, or side effects.
- Keep the question short and direct.
- Do not explain the code or provide answers.

Output:
Question: <your question here>
\end{promptbox}

\subsection{Question Extraction}

\textbf{Type:} Single user prompt \\
\textbf{Purpose:} Extract only the final question from generated text

\subsubsection*{User Prompt Template}

\begin{promptbox}[User Prompt]
Extract only the question from the following text.
Return the question exactly, with no extra words or labels:

{generated_text}
\end{promptbox}

\subsection{Answer Generation for Code Comprehension}

\textbf{Type:} System + User message \\
\textbf{Purpose:} Generate a detailed answer to a comprehension question

\begin{promptbox}[System Instructions]
You are an expert in evaluating code comprehension. The user will provide
code and a question about it.

Your goal is to generate one relevant answer in English.

The answer should focus on:
- Essential mechanisms of how the code works
- Important design decisions
- Potential pitfalls or unexpected behaviors

Provide a clear and thorough answer demonstrating deep understanding.
\end{promptbox}

\subsubsection*{User Message}

\begin{verbatim}
<JOINED CODE CHUNKS>

<Question about the code>
\end{verbatim}

\subsection{MCQ Answer Sanitization}

\textbf{Type:} Single user prompt \\
\textbf{Purpose:} Convert a verbose answer into a concise MCQ-style answer

\begin{promptbox}[User Prompt]
You are an expert Python developer and technical writer.

I will give you:
1. A Python code snippet
2. A question about that code
3. A detailed answer

Your task is to sanitize the answer:
- Remove fluff and redundancy
- Keep only what directly answers the question
- Make it short, clear, and direct
- Do not repeat the question
- Do not rephrase the code

Input Code:
{code}

Question:
{question}

Original Answer:
{answer}

Sanitized Answer:
\end{promptbox}

\subsection{Distractor Generation for MCQs}

\textbf{Type:} Single user prompt \\
\textbf{Purpose:} Generate exactly three plausible distractor answers for a programming MCQ

\begin{promptbox}[User Prompt]
You are an expert MCQ generator specializing in programming assessments.

Given the following:

Code:
{code}

Question:
{question}

Correct Answer:
{answer}

Generate exactly 3 plausible distractor answers (incorrect but believable options)
to be used in a multiple-choice question. Each distractor should:

1. Be contextually relevant to the code and question
2. Represent a different level of Bloom's Taxonomy (e.g., Understanding, Applying, Analyzing)
3. Be plausible -- choices a well-meaning but mistaken student might select
4. Be similar in structure or terminology to the correct answer
5. Avoid being trivially or obviously incorrect

Return ONLY the distractors as a valid Python list of dictionaries:

[
    {
        "option": "Distractor text here",
        "bloom_level": "Bloom's taxonomy level (e.g., Understanding, Applying, Analyzing)"
    },
    ...
]
\end{promptbox}

\subsection{Correct Answer Adaptation to Match Distractor Style}

\textbf{Type:} Single user prompt \\
\textbf{Purpose:} Rephrase the correct answer to match the style and structure of generated distractors

\begin{promptbox}[User Prompt]
You are an expert MCQ generator specializing in programming assessments.

Given the following:

Code:
{code}

Question:
{question}

Correct Answer:
{answer}

Here are 3 distractor answers generated for this question:
{distractor_examples}

Rephrase the correct answer so that it resembles the distractors in style, structure, 
and terminology, but remains fully correct.

Return ONLY the adapted answer as a string, with no extra explanation or formatting.
\end{promptbox}

\subsection{Multiple-Choice Question Prompt for Model Benchmarking}

\textbf{Type:} Single user prompt \\
\textbf{Purpose:} Ask a model to select the correct answer letter (A--D) for a code-related MCQ

\begin{promptbox}[User Prompt]
You are given a piece of code, a related question, and four multiple-choice options 
labeled A through D. Analyze the code, read the question carefully, and choose the 
correct answer by responding with only the letter (A, B, C, or D).

Code:
{code}

Question:
{question}

Options:
{formatted_options}

Answer (respond with A, B, C, or D only):
\end{promptbox}

\textit{Note:} The \texttt{formatted\_options} placeholder contains the code snippet, question text, and options dictionary formatted as shown in Section~\ref{sec:placeholders}.

\subsection{Answer Extraction Using Helper Model}

\textbf{Type:} System + User messages \\
\textbf{Purpose:} Extract only the correct answer letter from a model's response, optionally processing reasoning output

\begin{promptbox}[System Instructions]
Extract the letter corresponding to the correct answer from the following response.
The output must be only the letter, with no extra explanation or characters.
\end{promptbox}

\subsubsection*{User Message Sequence}

\begin{promptbox}[User Message]
Example conversation history:
1. User: "The answer to the question is A."
   Assistant: "A"
2. User: "B."
   Assistant: "B"
3. User: "C"
   Assistant: "C"

Followed by the actual response to extract:
{conclusion}
\end{promptbox}

\textit{Note:} The \texttt{conclusion} is extracted from the benchmark model output, with optional removal of \texttt{<think>...</think>} tags if reasoning extraction is enabled.

\subsection{Placeholders Explained}
\label{sec:placeholders}

\begin{description}
    \item[\texttt{\{entity\_name\}}] The name of the specific entity being analyzed
    \item[\texttt{\{entity\_A\}}, \texttt{\{entity\_B\}}] Names of two interacting entities
    \item[\texttt{\{entity\_A\_definition\_code\}}] Code defining entity A
    \item[\texttt{\{entity\_B\_definition\_code\}}] Code defining entity B
    \item[\texttt{\{entity\_interaction\_code\}}] Code showing how the entities interact
    \item[\texttt{\{generated\_text\}}] Text from which to extract a question
    \item[\texttt{\{code\}}] The source code snippet for the MCQ
    \item[\texttt{\{question\}}] The text of the multiple-choice question
    \item[\texttt{\{answer\}}] The correct answer to the question
    \item[\texttt{\{distractor\_examples\}}] The three generated distractor options
    \item[\texttt{\{formatted\_options\}}] Options A--D formatted as:
    \begin{verbatim}
    A. option_text
    B. option_text
    C. option_text
    D. option_text
    \end{verbatim}
    \item[\texttt{\{conclusion\}}] The benchmark model output after optional thought extraction
    \item[\texttt{\{llm\_output\}}] Raw model output potentially containing \texttt{<think>} tags
\end{description}

\textit{All placeholders are replaced with actual content before sending to the model.}

\end{document}